\documentclass[aps,prl,preprint,showpacs,amsmath,floatfix,superscriptaddress]{revtex4}

\usepackage{graphicx}

\begin{document}

\title{Quasiparticle band-edge energy and band offsets of monolayer of molybdenum and tungsten chalcogenides}

\author{Yufeng Liang}
\affiliation{Department of Physics, Washington University in St.
Louis, St. Louis, MO 63130, USA}

\author{Shouting Huang}
\affiliation{Department of Physics, Washington University in St.
Louis, St. Louis, MO 63130, USA}

\author{Ryan Soklaski}
\affiliation{Department of Physics, Washington University in St.
Louis, St. Louis, MO 63130, USA}

\author{Li Yang}
\affiliation{Department of Physics, Washington University in St.
Louis, St. Louis, MO 63130, USA}

\date{\today}

\begin{abstract}
We report the quasiparticle band-edge energy of monolayer of
molybdenum and tungsten dichalcogenides, MX$_2$ (M=Mo, W; X=S, Se,
Te). Beyond calculating bandgaps, we have achieved converged
absolute band-edge energies relative to the vacuum level. Compared
with the results from other approaches, the GW calculation reveals
substantially larger bandgaps and different absolute quasiparticle
energies because of enhanced many-electron effects. Interestingly,
our GW calculations ratify the band-gap-center approximation,
making it a convenient way to estimate band-edge energy. The
absolute band-edge energies and band offsets obtained in this work
are important for designing heterojunction devices and chemical
catalysts based on monolayer dichalcogenides.
\end{abstract}


\maketitle

Recently, two-dimensional (2D) semiconducting monolayer and
few-layer dichalcogenides have drawn significant interest from
researchers in light of the materials' exciting chemical,
electrical, and optical properties \cite{2005Novoselov, 2010heinz,
2010wang, 2011kis, 2008chen, 2013zhang, 2005norskov, 2007horch,
2011dai}. For example, enhanced spin-orbital coupling and unique
optical selection rules make these materials promising for
spintronics applications \cite{2012xiao, 2012feng, 2012heinz,
2012cui}. Accordingly, the electronic structure and, in
particular, the quasiparticle energy of 2D dichalcogenides have
been intensively studied to date. It is of particular interest
that first-principles GW calculations have shown that enhanced
many-electron effects dictate bandgaps of these 2D semiconductors
\cite{gw-1, gw-2}.

However, many important properties of semiconductors are not
solely decided by the bandgap. For instance, the relative
band-edge energies between different semiconductors and
corresponding band offsets are of fundamental interest in solid
state physics and are indispensable for the design of
heterojunction devices \cite{offset-1, offset-2, offset-3}.
Dichalcogenides have been known catalysts for years
\cite{2005norskov, 2007horch, 2011dai, 1996uemura}. Understanding
the ways in which quantum confinement modifies the absolute
band-edge energy and associated charge-transfer processes of
chemical reactions in these monolayer or few-layer semiconductors
is of critical importance for their catalytical applications.
Recently, substantial advances have been achieved in obtaining
qualitative band offsets \cite{2013wu, 2012jiang}, but there has
been extremely limited progress towards overcoming the bandgap
problem and including enhanced many-electron effects in order to
achieve accurate quasiparticle energies in monolayer
dichalcogenides .

Here we employ the well-established first-principles GW approach
to solve the aforementioned problems. Usually obtaining the
absolute band-edge energy and band offsets of semiconductors
requires, at least, two conditions: 1) a reference energy level
and 2) an accurate quasiparticle energy. Because we are
considering isolated samples of these 2D dichalcogenides, it is
natural to choose the surrounding vacuum as the reference energy.
The more sophisticated challenge, however, is obtaining the
quasiparticle energy. In particular, we must ensure that the
absolute energies are well-converged; this process is
significantly more costly than is the bandgap calculation
\cite{gw-4, gw-5}. In this vein, approximations that estimate the
absolute band-edge energy but avoid a fully-converged GW
calculation have been proposed \cite{2011carter} but their
validity has not yet been verified in dichalcogenides. Given this
context, the simple atomic structures and relatively inexpensive
cost of fully-converged GW calculations for monolayer
dichalcogenides make these systems excellent vehicles for
obtaining reliable absolute band-edge energies and, moreover,
ratifying the aforementioned approximations.

In this study, our calculation provides the quasiparticle energy
and corresponding band offsets of monolayer dichalcogenides via
the single-shot G$_0$W$_0$ calculation. The enhanced many-electron
interactions in such confined 2D semiconductors significantly
enlarge the bandgap and change the absolute band-edge energy
accordingly. The absolute band-edge energy and band offsets from
the GW calculation are substantially different from those of
density functional theory (DFT) and hybrid functional theory
(HFT). On the other hand, the types of band alignments from DFT
and HFT qualitatively agree with the GW results, meaning DFT and
HFT are valuable for band-alignment estimations, especially given
their inexpensive simulation costs. Interestingly, we find that
the band-gap-center model gives a surprisingly accurate absolute
band-edge energy without requiring a converged GW calculation.
Ultimately, the absolute quasiparticle band-edge energies and band
offsets obtained in this work will be helpful for designing
heterojunctions and catalysts comprised of these materials.

We apply the generalized gradient approximation (GGA) of
Perdew-Burke-Ernzerhof (PBE) as the exchange-correlation
functional in the DFT calculation \cite{pbe}. The single-shot
G$_0$W$_0$ calculation is employed to obtain the quasiparticle
energy. The spin-orbital coupling is not considered in this study.
The plane-wave cutoff for the DFT calculation is set to be 80 Rys.
The norm-conserving pseudopotentials \cite{1991Troullier} of
molybdenum and tungsten include the 4s4p and 5s5p semi-core
electrons, respectively. The k-point sampling is 12x12x1 for both
DFT and GW calculations. The dielectric-function energy cutoff is
set to be 10 Rys and the generalized plasmon-pole model (GPP) is
applied to obtain the dynamical screening \cite{louie-1}. A slab
Coulomb truncation is applied to mimic the isolated monolayer
structure with a vacuum spacing of 23 $\AA$ between adjacent
layers. All structures are fully relaxed according to the force
and stress by the DFT/PBE calculation. Their relaxed lattice
constants, listed in Table \ref{tb1}, are well consistent with
previous results \cite{gw-1, 2013wu}.

\begin{figure}
\includegraphics*[scale=0.30]{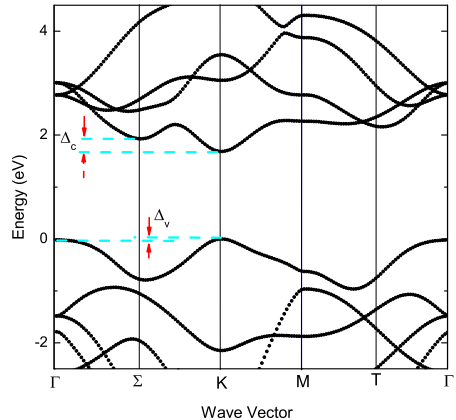}
\caption{(Color online) DFT-calculated band structure of monolayer
MoS2. The top of valence band is set to be zero. The energy
difference between the conduction band minimum at the K point and
the local minimum at the $\Sigma$ point is denoted by $\Delta_c$.
The energy difference between the valence band maximum at the K
point and the local maximum at the $\Gamma$ point is denoted by
$\Delta_v$.} \label{band-1}
\end{figure}

The general features of the band structures of studied monolayer
molybdenum and tungsten dichalcogenides are similar. As an
example, we plot the bandstructure of MoS$_{2}$, which exhibits a
direct bandgap, via DFT calculation in Fig. \ref{band-1}. As
revealed by many other works, there is another local maximum of
the valence band at the $\Gamma$ point, whose energy is very close
to the valence band maximum (VBM) at the K point. Interestingly,
this local maximum at the $\Gamma$ point will increase to become
the VBM in few-layer MoS$_{2}$ and thus the overall band structure
turns out to possess an indirect bandgap, which dramatically
changes the photoluminescence. \cite{2010wang}

\begin{table}
\caption{\label{tb1} Structure and electronic properties of
calculated monolayer dichalcogenides: lattice constant a, band gap
$E_g$, the energy difference $\Delta_v$ and $\Delta_c$ defined in
Fig. \ref{band-1}. (The HSE result is read from Reference 22.)}
\begin{center}
\begin{tabular}{ccccccccc}
\hline \hline

 & a (\AA)& $E_g^{DFT}$ (eV) & $E_g^{HSE}$ (eV) & $E_g^{GW}$ (eV) & $\Delta_v^{DFT}$ (eV) & $\Delta_v^{GW}$ (eV) & $\Delta_c^{DFT}$ (eV) & $\Delta_c^{GW}$ (eV) \\
\hline
 MoS$_2$ & 3.18 & 1.69 & 2.02 & 2.75 & 0.02 & 0.16 & 0.25 & 0.23 \\
 MoSe$_2$ & 3.31 & 1.43 & 1.72 & 2.33 & 0.23 & 0.34 & 0.23 & 0.33  \\
 MoTe$_2$ & 3.51 & 1.10 & 1.28 & 1.82 & 0.59 & 0.83 & 0.15 & 0.34 \\
 WS$_2$ & 3.20 & 1.78 & 1.98 & 2.88 & $<$0.002 & 0.06 & 0.25 & 0.25 \\
 WSe$_2$ & 3.33 & 1.50 & 1.63 & 2.38 & 0.26 & 0.34 & 0.21 & 0.36 \\
 WTe$_2$ & 3.52 & 1.10 & 1.03 & 1.77 & 0.65 & 0.79 & 0.45 & 0.39 \\
\hline \hline
\end{tabular}
\end{center}
\end{table}

In order to track the change of this subtle but important change
in band structure, we denote the energy difference as $\Delta_v$,
which is marked in Fig. \ref{band-1}, and list its values in Table
\ref{tb1}. For most monolayer dichalcogenides, $\Delta_v^{DFT}$ is
positive and larger than 20 meV. The only exception is WS$_2$,
whose $\Delta_v^{DFT}$ is almost zero. However, the further
inclusion of the spin-orbital coupling usually increases the VBM
at the K point \cite{2013wu}, preserving monolayer WS$_2$ as a
direct bandgap semiconductor. Moreover, from Table \ref{tb1}, we
can see that the value of $\Delta_v^{DFT}$ increases as the
group-VI element of dichalcogenides changes from S, Se, to Te.
Meanwhile, we have marked the energy difference, $\Delta_c$,
between the lowest conduction band at the K point, which is the
CBM, and that at the $\Sigma$ point. The corresponding data are
also listed in Table \ref{tb1}. At the DFT level, $\Delta_c$ is
around a few hundreds meVs for all studied monolayer
dichalcogenides.

Having applied the single-shot G$_0$W$_0$ approach to calculate
the quasiparticle energy of those six monolayer dichalcogenides,
the results are summarized in Table \ref{tb1}. First, the GW
correction significantly enlarges the bandgap for all studied
dichalcogenides. This enhanced many-electron effect has been
widely observed in many reduced dimensional semiconductors as a
result of depressed screening and stronger electron-electron
(\emph{e-e}) interactions \cite{catalin04, alouani06, yang07}. Our
GW calculated bandgaps are in good agreement with previous results
\cite{gw-1, gw-2}. Meanwhile, we have listed the bandgaps as
calculated by HFT with the HSE functional \cite{hse} read from
Ref. \cite{2013wu}. It can be seen that the GW bandgaps are
significantly larger than those from HFT/HSE.

From Table \ref{tb1}, one can see that the direct bandgap feature
is preserved for all of our calculated monolayer dichalcogenides:
the signs of all energy differences, $\Delta_v$ and $\Delta_c$ ,
remain unchanged after GW corrections. We find that the inclusion
of the 4s and 5s semi-core electrons is crucial for preserving the
direct bandgap feature in the GW calculation; otherwise, the local
minimum of the lowest conduction band at the $\Sigma$ point would
be the CBM, resulting in an indirect band gap. Our result is also
slightly different from another previous work, in which an
indirect bandgap of the WSe$_{2}$ structure is observed.
\cite{gw-1} This difference could be from the spin-orbital
coupling.

\begin{figure}
\includegraphics*[scale=0.33]{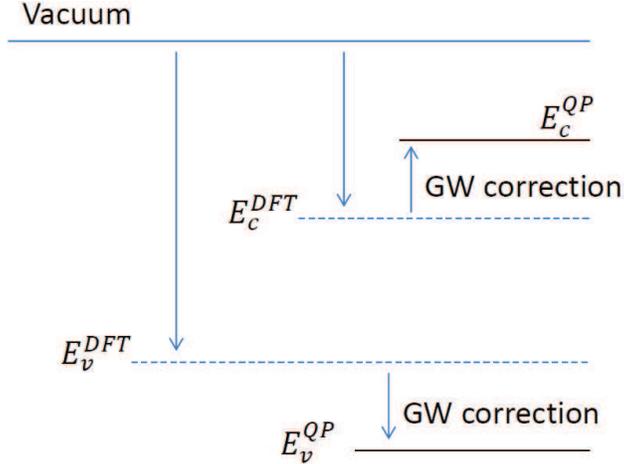}
\caption{(Color online) Schematic illustration of the absolute
band energy at the DFT and GW levels, respectively, relative to
the vacuum level.} \label{level}
\end{figure}

Beyond the quasiparticle bandgap, we have calculated the absolute
band-edge energy relative to the vacuum level \cite{2012jiang,
gw-5}. The absolute band-edge energy at the DFT level is referred
to the vacuum level that is defined by the potential energy in the
vacuum between dichalcogenide monolayers in our supercell
arrangement, as shown in Fig.~\ref{level}. Then we preform the GW
calculation and superpose the self-energy corrections to the DFT
eigenvalues, obtaining the absolute quasiparticle energy relative
to the vacuum level. In Fig.~\ref{level}, the final absolute
quasiparticle band-edge energies are $E^{QP}_c$ and $E^{QP}_v$ for
the CBM and VBM, respectively.

\begin{figure}
\includegraphics*[scale=0.30]{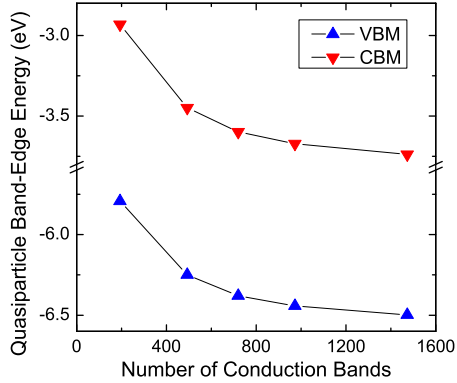}
\caption{(Color online) The convergence of the quasiparticle
energy of the CBM and the VBM, respectively, according to the
number of conduction band included in the GW calculation.}
\label{convg}
\end{figure}

Unlike the bandgap calculation, the convergence of the absolute
quasiparticle band-edge energy with respect to the number of
unoccupied conduction states included in the self-energy
calculation is very slow. For example, we present the convergence
of the CBM and VBM of monolayer MoS$_{2}$ in Fig. \ref{convg}.
Although the quasiparticle bandgap is reasonably converged at a
value of 2.75 eV after including around 200 conduction bands, the
absolute values of CBM and VBM do not reach their converged values
until we include around 1500 conduction bands.

In order to understand the slow convergence of the absolute
quasiparticle energy, we must examine the details of self-energy
in the GW calculation. Usually, the self-energy correction is
comprised of two contributions according to their physical
origins, the Coulomb-hole (COH) term and the screened-exchange
(SEX) term \cite{louie-1}. The aforementioned slow convergence is
mainly due to the COH term that involves the summation of an
infinite number of conduction bands, in principle \cite{louie-1}.
We find that the SEX term also converges slowly, although it is
faster than the COH term. Thus we use around 500 conduction bands
for the calculation of the static screening and around 1500
conduction bands for the final self-energy calculation.

\begin{figure}
\includegraphics*[scale=0.45]{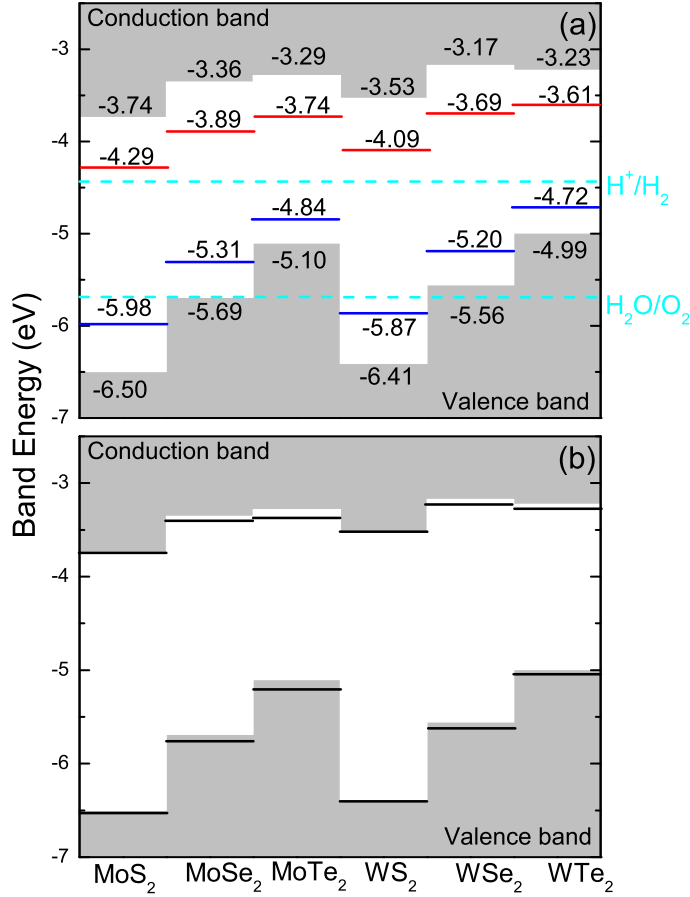}
\caption{(Color online) The absolute band-edge energy of
calculated monolayer dichalcogenides relative to the vacuum level
to the vacuum level. (a) The blue and red dashed lines stand for
the DFT/PBE results while the grey-shadow regions stand for the GW
results. The water reduction ($H^+/H_2$) and oxidation
($H_2O/O_2$) are marked by the cyan dashed lines, respectively.
(b) The absolute band-edge energies by the fully converged GW
simulation (grey-shadow regions) and the band-gap-center
approximation (solid dark lines). } \label{band-2}
\end{figure}

Finally, our calculated absolute quasiparticle band-edge energies
are summarized in Fig. \ref{band-2} (a), in which the DFT results
are also listed for reference. The enhanced many-electron
interactions in monolayer dichalcogenides substantially changes
the absolute band-edge energy from the DFT results. However, the
general trend of the evolution of the band-edge energies are
similar for both DFT and GW results. For instance, the band-edge
energy of MX$_2$ gradually increases as X varies from S to Te or M
varies from Mo to W. A particularly interesting point is that the
self-energy corrections modify both valence band and conduction
band-edge energies similarly, as seen from Fig. \ref{band-2} (a).
This is substantially different from the corrections found by
previous HFT studies, in which the corrections mainly affect the
VBM \cite{2013wu}.

The band alignments in Fig. \ref{band-2} exhibit several unusual
features. First, even after the costly GW calculation, except the
WSe$_2$/WTe$_2$ interface, the qualitative types of band
alignments for these materials from DFT and HFT/HSE have not
changed. For example, all of these calculations consistently
predict that the interface of MoS$_{2}$ and MoSe$_{2}$ has a type
II (staggered) band alignment. Secondly, the values of
GW-calculated band offsets are larger than those from DFT or HFT,
mainly due to larger bandgap corrections. Therefore, a
sophisticated calculation, such as the GW method, may be necessary
in order to obtain the quantitative band offset for
heterojunctions of 2D chalcogenides, while DFT or HFT calculation
can be convenient when assessing the type of the band alignment or
other properties \cite{hse0}. On the other hand, for
heterojunctions of our studied 2D chalcogenides with other
semiconductors, our calculated absolute band-edge energy shall be
crucial to decide the band offsets and even alignments.

Earlier work predicts that monolayer MoS$_2$ and WS$_2$ may work
for water splitting \cite{2013wu}. Hereby we have marked the
energy levels for the oxidation and reduction processes of water
splitting in Fig. \ref{band-2} (a). The GW calculation yields a
similar conclusion although the VBMs are usually lower and the
CBMs are generally higher than those of DFT and HFT results.

Previously, in order to avoid the slowly converging absolute band
energy, the band-gap-center approximation was proposed to estimate
the absolute band-edge energy with the assumption that the
self-energy correction shifts both CBM and VBM by similar amounts
but in inverse directions \cite{2012jiang, 2011carter}.
Interestingly, we find this model works very well for our studied
monolayer dichalcogenides. As shown in Fig. \ref{band-2} (b), the
band-gap-center approximation gives nearly the same band-edge
energy as the costly direct GW calculation. As discussed above,
the reason for this agreement is that our converged GW calculation
yields similar corrections for both the DFT-calculated VBM and
CBM. Therefore, this band-gap-center approximation may be
particularly useful for estimating the band offset of 2D
dichalcogenides because it only requires the quasiparticle
bandgap.

One must to be cautious when applying our absolute band-edge
energy towards realistic applications. Here we only consider the
isolated monolayer structures surrounded by vacuum. However, for
realistic conditions, the environmental effects will be extremely
important for these ultra-thin layer of semiconductors. For
example, the background dielectric response may substantially
reduce the self-energy corrections, affecting the band gap and
band offset dramatically \cite{gw-4, louie-2}. Moreover, for
photocatalytic processes, such as water splitting, excitonic
effects must be included since such processes are driven by
optically-excited excitons. In particular, electron-hole
(\emph{e-h}) interactions are known to be enhanced in monolayer
chalcogenides and in many other reduced-dimensional semiconductors
\cite{gw-1,yang-1}. Thus \emph{e-h} interactions will
substantially reduced the energy of the optical absorption edge,
making it significantly different from the quasiparticle bandgap.
In this sense, more sophisticated calculations including the
environment effects and the impact of excitons are desirable,
which is a major thrust in the field nanotechnologies as well.
However, our calculation serves as a valuable foundation for such
studies.

In conclusion, we have employed a first-principles GW calculation
to obtained the quasiparticle band structure and absolute
band-edge energy of monolayer dichalcogenides. Our converged GW
simulation not only produces the bandgaps but also provides the
band offsets of relevant heterojunctions. Both the bandgap and
absolute band-edge energy are substantially different from
previous DFT and HFT/HSE results. Surprisingly, the
band-gap-center model works very well for obtaining the absolute
band-edge energy without a fully-converged GW simulation, making
it a convenient way to estimate the band offsets and chemical
activity of monolayer dichalcogenides.

We acknowledge the support by NSF Grant No. DMR-1207141. The
computational resources have been provided by Lonestar of Teragrid
at the Texas Advanced Computing Center (TACC). The ground state
calculation is performed by the Quantum Espresso
\cite{2009Giannozzi}. The GW calculation is done with the
BerkeleyGW package\cite{2012Deslippe}.

\end{document}